\documentclass[aps,prb,twocolumn,showpacs]{revtex4}
\usepackage{graphicx}

\newcommand{\be}{\begin{equation}}
\newcommand{\ee}{\end{equation}}
\newcommand{\bea}{\begin{eqnarray}}
\newcommand{\eea}{\end{eqnarray}} 
\newcommand{\les}{\ell_{\hbox{\tiny ES}}}
\newcommand{\jes}{j_{\hbox{\tiny ES}}}
\newcommand{\ld}{\ell_{\hbox{\tiny D}}}
\newcommand{\tsec}{t_{\hbox{\tiny 2nd}}}
\newcommand{\tadv}{t_{\hbox{\tiny adv}}}

\newcommand{\pmax}{p_{\hbox{\tiny max}}}
\newcommand{\tins}{\tau_{\hbox{\tiny ins}}}
\newcommand{\tnucl}{\tau_{\hbox{\tiny nucl}}}
\newcommand{\tdim}{\tau_{\hbox{\tiny dim}}}

\newcommand{\e}{\epsilon}

\begin{document}

\title{Breakdown of metastable step-flow growth on vicinal
surfaces induced by nucleation}
\author{Daniele Vilone}
\affiliation{Dipartimento di Fisica, Universit\`a di
Roma ``La Sapienza'' and Center for Statistical Mechanics and Complexity,
INFM Unit\`a Roma 1, P.le A. Moro 2, 00185 Roma, Italy}
\author{Claudio Castellano}
\affiliation{Dipartimento di Fisica, Universit\`a di
Roma ``La Sapienza'' and Center for Statistical Mechanics and Complexity,
INFM Unit\`a Roma 1, P.le A. Moro 2, 00185 Roma, Italy}
\affiliation{Istituto dei Sistemi Complessi, CNR,
Via dei Taurini 9, 00185 Roma, Italy}
\author{Paolo Politi}
\affiliation{Istituto dei Sistemi Complessi, Consiglio Nazionale delle
Ricerche, Via Madonna del Piano 10, 50019 Sesto Fiorentino, Italy}
\date{\today}

\begin{abstract}
We consider the growth of a vicinal crystal surface in the presence
of a step-edge barrier. For any value of the barrier strength,
measured by the length $\les$,
nucleation of islands on terraces is always able to destroy
asymptotically step-flow growth.  The breakdown of the metastable
step-flow occurs through the formation of a mound of critical width
proportional to $L_c\sim 1/\sqrt{\les}$, the length associated to the
linear instability of a high-symmetry surface.
The time required for the destabilization grows exponentially with $L_c$.
Thermal detachment from steps or islands, or a steeper slope increase the
instability time but do not modify the above picture, nor change
$L_c$ significantly.  Standard continuum theories cannot be used
to evaluate the activation energy  of the critical mound and the
instability time.  The dynamics of a mound can be described as a one
dimensional random walk for its height $k$: attaining the critical
height (i.e. the critical size) means that the probability to grow
$(k\to k+1)$ becomes larger than the probability for the mound to shrink
$(k\to k-1)$. Thermal detachment induces correlations in the random
walk, otherwise absent.
\end{abstract}

\pacs{81.10.Aj, 68.55.Ac ,05.70.Ln}

\maketitle

\section{Introduction}
\label{sec_intro}

The growth process of a crystal surface by deposition of particles
resembles in many respects a phase separation process.
The role of the order parameter is played  by the local slope $u$,
whose average value $m$ is fixed by the orientation of the substrate where
deposition occurs. A uniform orientation is often not stable for a
growing crystal, for thermodynamic~\cite{ther_ins} as well as for
kinetic~\cite{review} reasons: As deposition proceeds
the flat profile gives way to a rough surface, with the formation and
growth of mounds.
The first stages of the destabilization process may occur through the
appearance of large wavelength undulations or the formation of
localized perturbations:
the former happens if the orientation $m$ is linearly unstable, the latter
if it is metastable. By varying $m$ it is therefore possible
to pass from one regime to another, much in the same way as it is possible
to go, in the phase separation of a binary alloy, from spinodal
decomposition to droplet nucleation
by changing the relative concentration.\cite{phase_sep}

In the context of crystal growth, the unstable regime corresponds to
a high-symmetry orientation, the counterpart of a symmetric quenching
for a binary alloy. This regime has been studied in
dozens of papers~\cite{high_symm} and similarities and differences 
with phase separation processes have been highlighted. In the following we
study the so called vicinal regime,\cite{vicinal} when the slope $m$ is
sufficiently far from a high symmetry orientation: in this case
the surface is made up of small terraces separated by steps and most
of the deposited atoms attach to preexistent steps (step-flow growth).
This regime is the counterpart of the case where, for a binary mixture,
the homogeneous phase is metastable, but it can be disrupted by the nucleation
and growth of droplets of the stable phase.

In the system considered in this paper, the key atomistic ingredient of the
destabilization is the presence of an energetic step-edge or
Ehrlich-Schwoebel (ES)
barrier~\cite{es} that hinders the descent of atoms from
upper to lower terraces. This instability has a purely kinetic
origin, due to the nonequilibrium character of the growth process.
The analytical investigation of a growing vicinal profile reveals two possible
linear instabilities: step-bunching~\cite{bunching} (the uniform density
of steps is unstable) and step-meandering~\cite{meandering} (straight steps
are unstable).
Step-edge barriers make the surface stable with
respect to the bunching and unstable with respect to the meandering. 
This picture does not exhaust all the possibilities, because mound
formation may occur also via nonlinear mechanisms,
leading to a late stage
morphology which does not differ from what is obtained by growing
on a (linearly unstable) high symmetry orientation. Two paths 
leading to mounding can be anticipated: atomistic nucleation on terraces
and defect formation, due to nonlinear meandering.~\cite{Kallunki04}
In the following we consider the first possibility, which
is the natural mechanism for two reasons: in the first place, it is solely
responsible for destabilization if steps keep straight;\cite{note_meander} 
secondly,
it is the correct process to consider if we want to study the
crossover from the unstable to the metastable regime,
because atomistic nucleation plus step-edge barriers are responsible
for destabilizing both a high symmetry and a vicinal surface.

Since our interest is the destabilization via atomistic nucleation,
we will consider a {\it one-dimensional} surface. This model may apply
to a two-dimensional surface when steps are straight 
(see the final section for additional remarks).

Let us now summarize the main features of the destabilization
mechanism and our principal results.
Freshly deposited atoms (adatoms) diffuse on terraces, generally
until they are incorporated at steps.
From time to time two of them meet and an atomistic nucleation occurs.
Since the vicinal flat profile is linearly stable, the nucleated dimer
is generally swept by the advancing steps. Only once in a while
another nucleation event takes place on top of the dimer before it is
reabsorbed, generating a very small mound.
A mound may grow larger, but in general it tends to disappear, unless,
by a fluctuation, it reaches a critical size such that it becomes
more likely for it to grow rather than to shrink.
At this point the flat profile is destabilized and it is not
recovered by the dynamics.
A crucial role is hence played by the critical nucleus,
the localized perturbation of the profile (mound) that separates small
perturbations that tend to be reabsorbed from large ones that grow
irreversibly.

Section~\ref{sec_crit-nuc} highlights the inadequacy of atomistic
approaches based on the assumption that the critical nucleus has a
fixed size, independent of the ES barrier.
Section~\ref{sec_cahn-hilliard} shows that also the application of
traditional continuum methods for studying the breakdown of
metastability fails in our case.
In Section~\ref{sec_j} we discuss the form of the unstable current,
which allows to distinguish between the unstable and the metastable 
regime and gives insight into the crossover connecting them.

In Secs.~\ref{sec_model} and~\ref{sec_ins-time} we present the
results of Kinetic Monte Carlo simulations showing that the surface
is metastable for any strength of the additional step-edge barrier
and that the instability time diverges exponentially for decreasing barrier.
We also show that when the slope is sufficiently small we have
a crossover to the linearly unstable regime, where the instability
time diverges as a power-law with the barrier.
Secs.~\ref{sec_mound-width} and \ref{sec_mound-height} present an
analysis of the properties of the critical nucleus,
whose width is related to the instability of a singular surface,
along with a detailed study of how the nucleus is dynamically generated,
allowing a quantitative derivation of the instability time.

Finally, in Sec.~\ref{sec_thermal} the global physical picture
is shown to be robust when thermal detachment is allowed: metastability
holds for any barrier strength also in this case.
Nonetheless, we show that thermal detachment has nontrivial effects
on the development of the critical nucleus and this results in a 
remarkable increase of the instability time.
A discussion (Sec.~\ref{sec_conc}) concludes the paper.

In a recent Letter,\cite{Vilone05} we have presented some partial
results of this line of research.
Though some overlap is unavoidable, we have minimized it in the present
paper by referring to the Letter when needed and focusing here 
on completely new results: the failure of the continuum approach
(Sec.~\ref{sec_cahn-hilliard}); the study of the crossover between
singular and vicinal regimes (Sec.~\ref{sec_j}); 
the detailed numerical and analytical investigation of the
destabilization process at the atomistic level (Secs.~\ref{sec_mound-width},%
\ref{sec_mound-height} and the Appendix); 
the simulations with  thermal detachment (Sec.~\ref{sec_thermal}).

\section{Theory with a critical nucleus of fixed size}
\label{sec_crit-nuc}

A simple and appealing argument for analyzing the destabilization
of step-flow on vicinal surfaces has been recently introduced by
Kallunki and Krug~\cite{Kallunkitesi,Kallunki04} and it
is based on the idea that the critical nucleus is a
mound of height two.
The stability or instability of step-flow is therefore assessed by
evaluating whether such a nucleus is formed or not by the dynamics.
This is performed by comparing the mean timescales of the two
processes that can occur once an island is formed on a vicinal terrace:
either the island is reabsorbed by the advancing step (and this requires
a time $\tadv$) or a second-layer nucleation event takes place on top of it
(this occurs over a time $\tsec$).
If $\tsec < \tadv$ a mound of height two is formed and this leads to
the destabilization of the surface.

There are several ways of implementing in practice this criterion and
they all lead to results that are equivalent, apart from small
differences in the numerical factors.
We refer here to the simplest model of epitaxial growth, with
deposition occurring at rate $F$, adatom diffusion at rate $D$ and
the strength of the ES barrier quantified by the length $\les=
D/D'-1$, where $D'$ is the interlayer diffusion coefficient.
One way to implement the criterion is to impose that the mean number
of second-layer nucleation events occurring during the deposition
of a monolayer is larger than 1.
The number of such nucleation events can be evaluated by integrating
the second-layer nucleation rate $\omega$ over time up to $1/F$,
\be
N_{nuc}(t=1/F) = \int_0^{1/F} ds \, \omega[R(s)],
\ee
where $R(s)$ is the size of the island at time $s$
and
\be
\omega(R) = \kappa {F^2 R^4 \over 12 D} \left(1+ 6 {\les \over R} \right),
\ee
($\kappa$ being a constant of the order of one,\cite{Elkinani94, Politi03}
that will be neglected in the following).

Assuming\cite{note_KK} that half of the deposited adatoms contribute to the
growth of the island (and the other half is incorporated at the advancing
steps), we have $R(s) = F \ell s/2$, where $\ell=1/m$ is the size of the
vicinal terrace. Performing the integrals and imposing
$N_{nuc}(t=1/F)>1$, we obtain
\be
\les > \les^c = 64 {D/F \over \ell^3} - {1 \over 15} \ell.
\label{lesc}
\ee
We stress again that
other implementations of the same criterion yield virtually identical results.
For instance one can impose that the probability $p_{nuc}(t^*)$
that second-layer nucleation occurs within a time $t^*=1/(2F)$
after the creation of the island is larger than $1/2$. In this
way we obtain the same condition~(\ref{lesc}) except for
a factor $\ln(2)$ multiplying the first term on the r.h.s..

We conclude that the assumption of a critical nucleus of height two implies
the existence of a threshold: for small enough barrier,
step-flow growth should be fully
stable with respect to the nucleation of mounds.
As shown below, this prediction is in striking contrast with numerical
simulations, showing that, even for values of $\les$ three
orders of magnitude smaller than $\les^c$, the surface is in fact metastable.
This is a clear indication that something in the argument just presented
is not correct. 

One could think that a higher critical nucleus would fix the problem.
However, a critical nucleus of fixed height $k_c$ would always
imply the existence of a range of small values of $\les$ for which
step-flow growth is fully stable. This can be seen by considering
that, for any $k_c$, the condition for instability is of the form
$A(\les) > A_0$, where $A$ is a growing function of $\les$, expressing
how likely the formation of the critical nucleus is,
and $A_0$ is a constant, whose precise value depends somewhat arbitrarily
on the details of the implementation.
If $A(\les=0)>A_0$ then the growth would be unstable even for $\les=0$ and
this is absurd. Hence $A(\les=0)$ must be smaller than $A_0$ and this
implies that in an interval of small $\les$ growth is stable.
Metastability down to $\les=0$ would be possible only if $A(\les=0)$
exactly equals $A_0$, but this is impossible, given that a precise
value of $A_0$ cannot be unambiguously defined.

Another reason for the failure of the atomistic argument discussed above
could be that comparing the average values of $\tsec$ and $\tadv$ is not
appropriate, we should consider instead their full distributions $P_1(\tadv)$
and $P_2(\tsec)$.
The surface is unstable if there is a finite probability
that $\tsec < \tadv$, i.e. if 
\be
p_{\tsec<\tadv} = \int_0^\infty d\tsec P_2(\tsec)\int_{\tsec}^\infty
d\tadv P_1(\tadv) > 0.
\label{eq_p12}
\ee
Let us now consider the limit $\les\to 0$:
both distributions have a well defined limit, because the nucleation
rate on top of a terrace and the velocity of a step reach
finite values for $\les=0$. Since $P_2(\tsec)>0$ for
small but finite $\tsec$, the quantity $p_{\tsec<\tadv}$
is always greater than zero. In simple terms, considering
the condition~(\ref{eq_p12}) for metastability leads to the wrong conclusion
that the vicinal surface should be destabilized for $\les=0$ as well.
In conclusion, the hypothesis of a critical nucleus with fixed height
is not compatible with numerical results and known properties for $\les=0$.
As it will be shown below, the diverging size of the critical nucleus
is the crucial element of the breakdown of metastable step-flow growth.

\section{Continuum Cahn-Hilliard theory for the metastability}
\label{sec_cahn-hilliard}

The goal of this section is to study the metastability by
applying the Cahn-Hilliard (CH) theory developed for ordinary phase-separation
to a minimal continuum model for epitaxial growth. We will show that
this approach is completely unable to explain the dynamical
behavior of the growing surface even in qualitative terms.
A very short introduction to the continuum approach is given in the
next two paragraphs.

In the absence of events which do not conserve matter (evaporation)
or volume (formation of overhangs), the dynamics of a one-dimensional
growing crystal
surface is described by $\partial_t z = - \partial_x J$,
where $J$ is the current
describing all microscopic processes occurring at the surface and
$z$ is the local height $h$ with respect to a comoving frame,
$z=h-Ft$ ($F$ is the flux intensity). 
Since we are not studying phenomena depending on the substrate-adsorbate
interaction, $J$
can be safely assumed to depend only on spatial derivatives of $z$,
$u=z', u', u'',\dots$. In the next section we will evaluate
numerically the slope dependent current, $\jes(u)$, also called
step-edge or ES current because it is due to asymmetric sticking to
steps. 
$\jes(u)$ is linear at small slope,\cite{review} attains a maximum
for $u=m_0$ and finally decreases. In addition, other terms
depending on higher order derivatives contribute to the current $J$.
The most important
has the form $Ku''$, is called Mullins term, and is due to several
microscopic processes:\cite{PV} thermal detachment from steps and the
random character of nucleation and diffusion. In the following we will consider
a minimal model, with $J=Ku'' + \alpha \jes(u)$. The prefactor $\alpha$
is chosen so that $\jes'(0)=1$: it might be included in the definition
of $\jes(u)$, but its explicit appearance will be useful later.

Given the form of $J$, the evolution equation for the surface slope $u$
has the form of a generalized CH equation
\be
\partial_t u = -\partial_x^2 [Ku''(x)+\alpha \jes(u)].
\label{eq_min}
\ee
Linear stability analysis of a vicinal surface of slope $m$, $z=mx + 
\delta z\exp(\omega t + iqx)$ provides the spectrum $\omega(q)=
\alpha \jes'(m) q^2 -Kq^4$. If $m>m_0$, $\jes'(m)<0$ and the surface is
linearly stable; if $m<m_0$ the surface is linearly unstable, the most
unstable wavevector (the one for which $\omega(q)$ is maximum) is
$q_c=\sqrt{\alpha \jes'(m)/2K}$, the most unstable length is
$L_c=2\pi/q_c=2\pi\sqrt{2K/\alpha \jes'(m)}$, and the linear
instability emerges after a time $t_c\approx 1/\omega(q_c)
\approx K/(\alpha \jes'(m))^2$. 
For small barriers~\cite{review} $\alpha$ is proportional to $\les$;
therefore the instability appears
with a typical lengthscale of order $1/\sqrt{\les}$ and after a
typical time of order $1/\les^2$. These temporal and spatial scales
also diverge when $m\to m_0^-$, because $\jes'(m)\to 0$.

Stationary solutions of Eq.~(\ref{eq_min}) are given by
$ Ku''(x) +\alpha \jes(u) = \mu $,
where $\mu$ is related to $m$ by the condition $\langle u\rangle=m$.
The critical nucleus is a localized steady state, characterized by
$u(x)=m$ everywhere except in a region of finite size. This implies
that, for the critical nucleus, $\mu= \alpha \jes(m)$ and
\be
Ku''(x) +\alpha[ \jes(u) -\jes(m)] = 0 .
\label{eq_nc}
\ee

Rather than reproducing the standard procedure for getting the energy
$\Delta {\cal F}$ of the critical nucleus~\cite{Krug95}, we simply
remark that rescaling $x$ allows to find how $\Delta {\cal F}$  depends
on $\alpha$
and $K$. If $X=\sqrt{\alpha\over K}x$, Eq.~(\ref{eq_nc}) has the
parameter-free form $u_{XX} +[\jes(u) -\jes(m)] = 0$ and the pseudo
free-energy ${\cal F}$ [such that
$\partial_t u=\partial_x^2\left( {\delta{\cal F} \over\delta u}\right)$] is
\bea
{\cal F} &=& \int dx [\frac{K}{2}(\partial_x u)^2 + \alpha U(u)]
\nonumber \\
&=&
\sqrt{\alpha K} \int dX [\frac{1}{2}(\partial_X u)^2 + U(u)]
\eea
where $U'(u)=-\jes(u)$. In conclusion,
\be
\Delta {\cal F} (\alpha,K) = \sqrt{\alpha K} \Delta {\cal F} (1,1).
\label{eq_e-nc}
\ee

In a thermodynamic system, once the activation energy 
$\Delta {\cal F}$ has been calculated, the instability time 
is given by
$\tins \sim \exp(\Delta {\cal F}/k_B T)$, because the
system overcomes the barrier by thermal fluctuations.
In the far-from-equilibrium case of a growing surface, fluctuations
allowing to system to leave the local minimum have different origins:
fluctuations in the flux (shot-noise) and fluctuations in the
nucleation and diffusion processes.
Their amplitude plays the role of $k_B T$ in the expression for $\tins$.

The analytical form of $\jes(u)$, valid for any $u$ and $\les$, is not
easy to write,
but we need here only the limiting form for vanishing barriers
and finite slope. With these caveats, all the $\les$ dependence is contained
in the prefactor $\alpha$, which is proportional to $\les$.
Hence for fixed slope, the unstable current is linear in the ES length
in the limit of small barrier. According to Eq.~(\ref{eq_e-nc}),
the activation barrier is therefore an {\it increasing}
function of $\les$, the opposite of what is expected and observed!
This surprising result cannot be changed by taking into account
the possible effects of $\les$ on the noise amplitude, because 
nucleation noise has a finite limit for $\les=0$.
We must conclude that the continuum minimal model, Eq.~(\ref{eq_min}),
is not appropriate for the study of the breakdown of step-flow growth.
Additional comments are deferred to the Conclusions.

\section{Crossover from singular to vicinal for a tilted surface}
\label{sec_j}

Before reporting the results of a numerical study of the breakdown
of step-flow growth, let us discuss what occurs at the boundary
between singular and vicinal surface orientations.
Linear stability analysis predicts two distinct types of dynamical
behavior, depending on how large is the average global tilt $m$ with respect
to the value $m_0$ for which the Ehrlich-Schwoebel nonequilibrium current
$\jes$ is maximal.
For $m<m_0$ (singular surfaces) a flat profile is linearly unstable for
$q<q_c$ (see the previous section).
For $m>m_0$ instead (vicinal surfaces),
perturbations of small amplitude decay, since the flat profile is linearly
stable, and, if any destabilization occurs, this has to be triggered by
fluctuations of sufficiently large amplitude.

In order to study the way the destabilization of vicinal surfaces takes
place, it is therefore crucial to check that the surface considered
is effectively vicinal, i.e., its slope is larger than $m_0$.
This point is made complicated by the fact that $m_0$ may in principle
(and it actually does) depend on $\les$,
so that it may happen that a surface with fixed $m$ is singular
or vicinal depending on the value of $\les$.
The variation of $m_0$ is not expected to be large, because
the form of the current predicts that the ES current is maximal
for finite slopes of the order of $1/\ld$ (see Ref.~\onlinecite{review})
for both $\les \to 0$ and
$\les \to \infty$ ($m_0^0$ and $m_0^\infty$, respectively).
In any case, this effect must be taken into account for a detailed
understanding of the destabilization process.
For this reason we have numerically evaluated the form of the nonequilibrium
current $\jes(u)$, by measuring, for the KMC model described below,
the imbalance in the number of hops toward left or right that adatoms
take during the
growth of the first few monolayers.\cite{Krug93}
More precisely, if $N_l$ and $N_r$ are the number of steps toward left
and right, respectively, then the current is
\be
\jes = {|N_l - N_r|\over N_{ad}},
\ee
where $N_{ad}$ is the number of adatoms deposited.

\begin{figure}
\includegraphics[angle=0,width=8cm,clip]{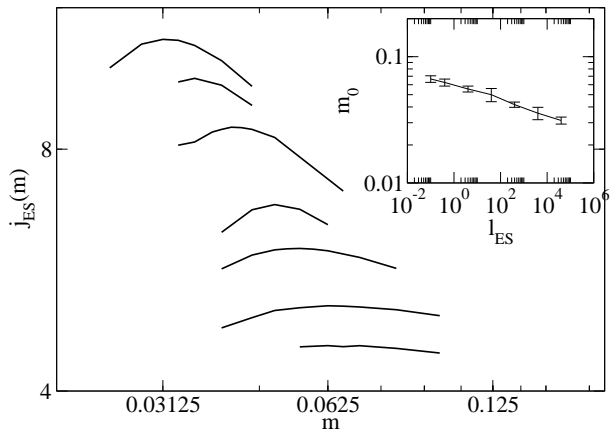}
\caption{Main: Plot of the nonequilibrium current $\jes(u)$ as a
function of the slope $m$,
for several values of the Ehrlich-Schwoebel length.
From bottom to top: $\les=0.1, 0.4, 4, 40, 400, 4000, 40000$.
Curves have been shifted along the vertical axis for clarity.
Inset: Plot of the value $m_0$ of the slope for which the
current is maximal.}
\label{Fig_jes}
\end{figure}

The results, reported in Fig.~\ref{Fig_jes}, display a very slow
crossover between $m_0 \approx 0.07$ for small $\les$ and
$m_0 \approx 0.03$ for the largest value of $\les$ we could
consider: we stress that
$m_0$ varies by a factor two against a variation of $\les$ over
five orders of magnitude.
As expected, these values are of the order of $1/\ld \approx 0.025$.

\begin{figure}
\includegraphics[angle=0,width=8cm,clip]{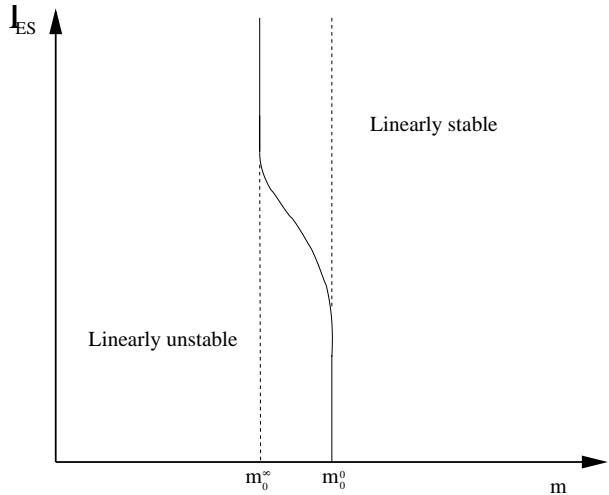}
\caption{Schematic plot of the different regimes as a function of the slope
$m$ and of the ES barrier.}
\label{Fig_fase}
\end{figure}

We can conclude that three types of behavior exist for the system
under consideration (Fig~\ref{Fig_fase}).
For $m<m_0^\infty \approx 0.03$ the surface is singular (hence linearly
unstable) for all values of the ES barrier.
For $m> m_0^0 \approx 0.07$ the surface is always linearly
stable and metastability can be studied down to $\les \to 0$.
For intermediate slopes the surface is vicinal for large $\les$, while
it becomes linearly unstable as $\les$ is decreased. In this case
a crossover is expected in the way the instability time depends on $\les$.

Notice that the value $m=1/15 \approx 0.067$ considered in the following
sections just falls within this interval.
It is important to remark (Sec.~\ref{sec_cahn-hilliard})
that the linear instability for a singular surface
does not imply that the instability develops immediately.
The time required for a sizeable amplification of the initial fluctuations
depends on $m$ and it {\em diverges} for $m \to m_0$ or $\les\to 0$.

\section{Kinetic Monte Carlo simulations}
\label{sec_kmc}

\subsection{The model}
\label{sec_model}

We perform Kinetic Monte Carlo (KMC) simulations of the simplest model
for epitaxial growth in $d=1$.
The surface is represented by a set of integer height variables $h_i$, 
defined on a one-dimensional lattice of $L$ sites.
An average tilt $m$ is imposed through helical boundary conditions,
$ h_{i+L}=h_i+m L$.
The average terrace size is thus $\ell=1/m$.
The initial condition is a regular train of steps,
$ h_i=\left[\frac{i}{\ell}\right]$, where $[x]$ is the integer part of $x$.
Deposition $(h_i\rightarrow h_i+1)$ occurs at rate $F$ on randomly selected
sites.
Singly bonded atoms (adatoms) attempt diffusion hops to nearest neighbor
sites, at rate $D$ if the neighbor belongs to the same terrace of the
first site, at rate $D'<D$ (because of the Ehrlich-Schwoebel effect)
if the atom must descend a step.
Dimers and larger islands are immobile.
Thermal detachment is possible at rate $R_d$. 
When a particle is detached from a step or an island
it is placed on the lower terrace at distance 1 from the step.
We have checked that if detachment events into the upper terrace
are also allowed (with the same rate of the detachments into the
lower terrace), results do not change appreciably.
The dynamics has been implemented via a rejection-free type of
algorithm.\cite{Bortz75,Barkemabook}

We have always kept fixed the deposition rate $F=1$ and the diffusion
coefficient $D = 5 \times 10^5$, so that the diffusion length
is $\ld \approx 40$, as in Ref.~\onlinecite{Krug95}.
The size of the system has been always $L=1000$, large enough 
($L \gg \Lambda_c$, see below) to allow for unstable mounds to form.
We have varied $D'$, in order to change the ES length $\les=D/D'-1$,
and $R_d$.

\subsection{The instability time}
\label{sec_ins-time}

In order to analyze the breakdown of metastability, we focus first
on the destabilization time $\tins$ , i.e. the time needed for the
formation of the first critical nucleus. The identification of this object
is not trivial and it has been discussed in Ref.~\onlinecite{Vilone05}.
Using the same numerical procedure for the identification of
the nucleus, we have computed $\tins$ for several values of the
slope $m$, of the rate for thermal detachment $R_d$, and with varying $\les$.

\begin{figure}
\includegraphics[angle=0,width=8cm,clip]{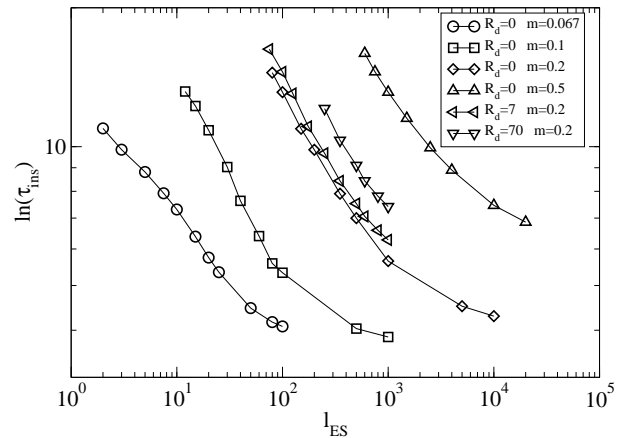}
\caption{Double logarithmic plot of $\ln(\tins)$ as a function of $\les$,
for several values of the slope $m$ and of the thermal detachment rate $R_d$.
Error bars are smaller than symbol size.}
\label{Fig_tauins}
\end{figure}

The double-logarithmic plot of $\ln(\tins)$ vs $\les$ (Fig.~\ref{Fig_tauins})
indicates a divergence as $\les \to 0$, with no clear-cut
exponential or power-law behavior.
Common (and intuitively expected) features are that increasing $m$ induces a
delay in the destabilization process, just as, for fixed $m$, thermal
detachment does.

\begin{figure}
\includegraphics[angle=0,width=8cm,clip]{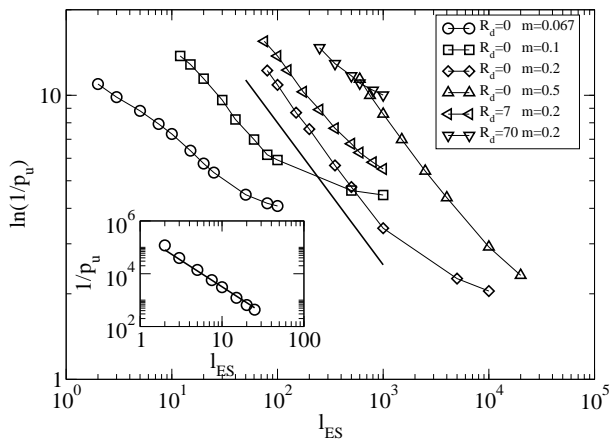}
\caption{Main: Double logarithmic plot of $\ln(1/p_u)$ as a function of $\les$,
for several values of the slope $m$ and of the thermal detachment rate $R_d$.
Error bars are smaller than symbol size.
A straight line indicates a divergence as $\exp(a/\sqrt{\les})$.
Inset: $1/p_u$ for $m=0.067$ and $R_d=0$ is plotted in a double
logarithmic-plot, showing that the
divergence for $\les \to 0$ is close to the $1/\les^2$ behavior (solid line)
predicted for a singular surface.}
\label{Fig_pu}
\end{figure}

More insight into how the destabilization process depends on $\les$
is obtained by considering the formation of the critical nucleus as the
combination of two distinct processes.
In the first place a dimer must be nucleated on a vicinal terrace.
Then the dimer must grow up to a mound of a certain critical size that
makes it unstable.
The first event occurs at a rate $1/\tnucl(L)$, which is proportional
to the system size $L$ and is independent of the ES
length in the limit $\les \to 0$.
The second event occurs only with a small probability $p_u$, strongly
dependent on $\les$ but not on $L$.
Hence we can write
\be
\tins(L,\les) = \tnucl(L,\les) {1 \over p_u(\les)}.
\label{eq_nucl-pu}
\ee
The value of $\tnucl$ can be written in its turn as
$\tdim(\ell,\les) \ell/L$
where $\tdim$ is the time to nucleate a dimer on a vicinal 
terrace of size $\ell$ and $L/\ell$ is the number of such terraces:
$\tdim$ could be computed analytically following
Ref.~\onlinecite{Politi03}, but it is straightforward to
evaluate it numerically. Furthermore, an accurate determination
of $\tnucl$ is important in order to evaluate correctly $1/p_u(\les)$,
which can be extracted by
plotting $\tins/\tnucl$ vs $\les$ (Fig.~\ref{Fig_pu}).

It turns out that, in almost all cases, $1/p_u$ diverges with $\les \to 0$
as $\exp(a/\les^\gamma)$, with $\gamma \approx 0.5$ and without evidence of
a finite threshold.
The only exception is the case of weaker slope, $m=0.067$,
where the divergence for small barrier is slower than exponential.
This different behavior can be understood on the basis of the form of the
unstable current $\jes$ reported in Fig.~\ref{Fig_jes}.
For $\les \le 10$ the maximum $m_0$ of the current is very close to the
slope $m=0.067$ considered here. As a consequence, as the barrier is reduced
the surface is more and more singular, rather than vicinal.
The destabilization takes place because of a (weak) linear instability
instead than through the breakdown of a metastable state. As a consequence
the dependence of $1/p_u$ on $\les$ is not exponential, as for the
other curves in Fig.~\ref{Fig_pu}.
The divergence is a power-law, with an effective exponent $2.2$
(see Fig.~\ref{Fig_pu}, inset), not far from the behavior $1/\les^2$
expected from the linear instability of a singular surface.

\subsection{The mound width and its critical value}
\label{sec_mound-width}

The evolution that, starting from a dimer nucleated on a
vicinal terrace, leads to the formation of an unstable mound,
is a very complicated process depending
at each time on the detailed form of the profile.
In order to understand it theoretically, one must focus on few
parameters.
The features that most naturally characterize a mound are its height
$k$ and its width $\Lambda$.
Both are useful in the analysis of the destabilization process.
We first discuss 
the role of the mound width, which, despite being numerically harder
to determine, it is conceptually more fundamental.
In the following subsection we will consider the temporal evolution of
the height, which is easily measured in simulations.

\begin{figure}
\includegraphics[angle=0,width=8cm,clip]{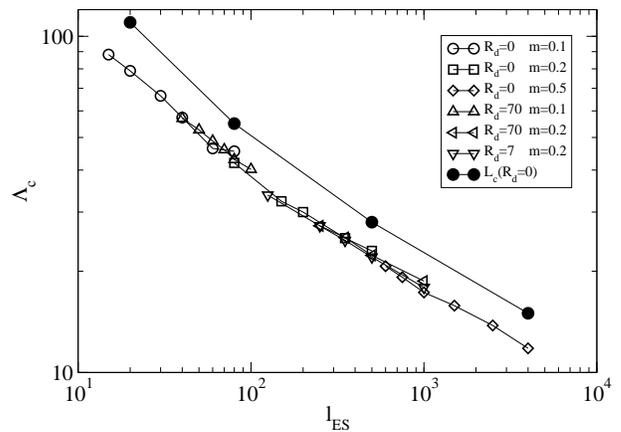}
\caption{Plot of the width $\Lambda_c$ of the critical mound 
as a function of the ES length for several values of the slope $m$
and the thermal detachment rate $R_d$. Error bars are smaller than
symbol size.
The full circles are the values of the length $L_c$ associated to the
linear instability for a singular surface.}
\label{Fig_Lambda}
\end{figure}

As discussed in Ref.~\onlinecite{Vilone05}, a good definition of the
mound width is the size of the terrace immediately below the top one.
For each value of $m, \les$ and $R_d$
it is thus possible to measure the critical width $\Lambda_c$, as the
average width of mounds that start to diverge irreversibly.
A plot of this quantity (Fig.~\ref{Fig_Lambda})
reveals the remarkable property that $\Lambda_c$ does not depend on $m$,
nor on the rate of thermal detachment $R_d$. Moreover the dependence
on the ES length is a power-law, with an effective exponent not far from 1/2.
We now focus on the case without thermal detachment leaving
for the next section the analysis of the effect of a finite $R_d$.

For the case without thermal detachment,
the behavior displayed in Fig.~\ref{Fig_Lambda} is naturally and consistently
interpreted~\cite{Vilone05} by relating the critical width
$\Lambda_c(\les)$ to the critical size $L_c(\les)$ associated to the
linear instability of a singular surface
\be
\Lambda_c(\les) \propto L_c(\les).
\label{Lambda_c}
\ee

This interpretation is further corroborated by a direct numerical
evaluation of $L_c(\les)$ as the smallest system size such that
for a singular surface the instability can take place.
The comparison of $\Lambda_c$ with $L_c(\les)$ (Fig.~\ref{Fig_Lambda})
strongly supports the validity of Eq.~(\ref{Lambda_c}), confirming that
a growing mound becomes unstable when its width is large enough to
trigger the same type of instability at work for a singular surface.
In this respect the width $\Lambda$ is the fundamental
quantity determining whether a mound is stable or unstable.
Because of the difficulties in its numerical determination~\cite{Vilone05},
it is not easy to analyze the detailed evolution of $\Lambda$ in order
to understand how the critical nucleus is generated by the dynamics.
In particular, we aim at deriving a quantitative formula for the
instability time $\tins$ or the relevant factor $1/p_u$.
For this purpose it is more useful to study the mound height $k$,
which is easily measured in simulations.

\subsection{Evolution of the mound height}
\label{sec_mound-height}

The height is not as fundamental as the width for the determination
of whether a mound is supercritical or not.
For example a critical height cannot in principle be rigorously defined.
This can be seen by taking a mound generated by the dynamics
just a bit wider than $\Lambda_c$ and strongly reducing its height:
it remains supercritical.
Nevertheless the dynamics couples the mound width and height so that
the evolution of $k$ tipically reflects that of $\Lambda$ and
a height $k_c$ of the critical nucleus can be in practice defined.

\begin{figure}
\includegraphics[angle=0,width=8cm,clip]{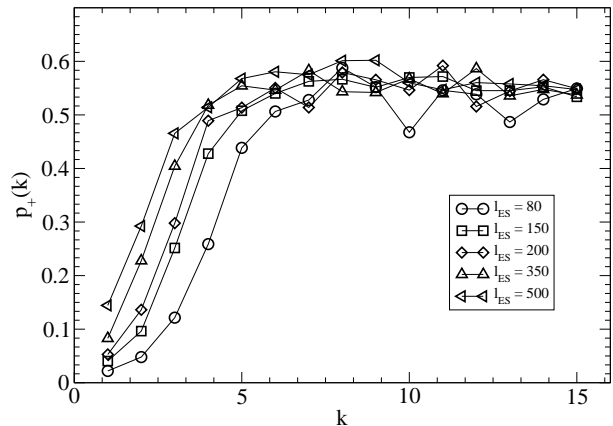}
\caption{The probability $p_+(k)$ that a mound of height $k$
eventually turns into
a mound of height $k+1$, rather than decaying to a mound of height $k-1$.
Results are for $m=0.2$ and $R_d=0$.}
\label{Fig_pofk}
\end{figure}

If a mound is described only in terms of its height, its dynamics
is reduced to a one-dimensional random walk (RW), with only two
possible processes for a mound of height $k$: it may either become taller
($k \to k+1$) via dimer nucleation on its top terrace, or shorter
($k \to k-1$) if the bottom terrace delimiting the mound is filled.
The increase (decrease) happens with probability $p_+(k)$ [$1-p_+(k)$].
It is possible to determine, via KMC simulations, the function $p_+(k)$,
which is reported in Fig.~\ref{Fig_pofk}.

The stochastic nature of destabilization for large $k$ is evident,
consequence of the shape of $p_+(k)$.
For small $k$, the probability that the mound
grows is much smaller than $1/2$. This implies that only rare events
lead to mound growth. This effect is reduced as $k$ is increased, up to
a certain value $k_c$, such that $p_+(k)>1/2$ for $k>k_c$.
The threshold $k_c$ plays the role of the effective height
of the critical nucleus.
The evolution is therefore biased toward the flat profile for $k<k_c$
while the bias is toward higher $k$ (destabilization) for $k>k_c$.
This translates into microscopic terms the metastable nature of a
vicinal surface:
it is stable with respect to small fluctuations, but unstable when a large
fluctuation by chance appears.
Fig.~\ref{Fig_pofk} shows that $k_c$ grows as $\les$ is reduced,
invalidating the assumption of the argument in Sec.~\ref{sec_crit-nuc}.

\begin{figure}
\includegraphics[angle=0,width=8cm,clip]{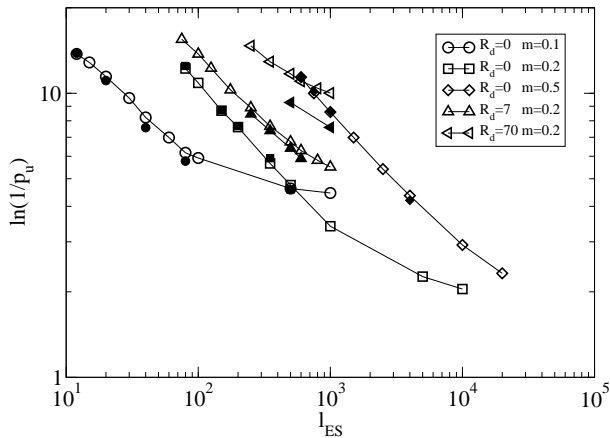}
\caption{Comparison between the values of $1/p_u$ obtained in numerical
simulations (empty symbols) and the same quantity computed within the
uncorrelated random walk with transition probabilities $p_+(k)$
(filled symbols).}
\label{Fig_pu_comp} 
\end{figure}

Using the values of $p_+(k)$ extracted from the full KMC simulations,
and taking them as the transition probabilities for an {\em uncorrelated}
one-dimensional random walk, it is possible to compute $1/p_u$ as the
fraction of times a walker starting in $k=1$ reaches a large reference
value (larger than $k_c$, say 15) without
touching the absorbing  boundary $k=0$. The results, presented in
Fig.~\ref{Fig_pu_comp}, agree very well with the outcome of simulations,
indicating that, despite its rather crude simplifications, the random
walk picture captures much of the destabilization process,
and it allows to recover the dependence of the instability time
$\tins$ on $\les$. 
It is important to remark that this result is not at all trivial.
Correlations in the random walk can generate, with the same transition
probabilities $p_+(k)$, very different values of $1/p_u$.
The agreement exhibited in Fig.~\ref{Fig_pu_comp} means that the uncorrelation
of RW steps is a correct assumption.
As it will be shown below, this is no more true when thermal detachment
is allowed.

Within the RW picture it is also possible to derive analytically a
formula for $1/p_u$. As shown in the Appendix one has to solve a
stationary convection-diffusion equation with a space-dependent drift related
to the form of $p_+(k)$.
By looking at Fig.~\ref{Fig_pofk} it is reasonable to take the explicit
espression $p_+(k)=p_0 e^k$ for $k<k_c$ and $p_+(k)=\pmax$ for $k>k_c$,
with $p_0 e^{k_c} = \pmax$. In this way we obtain (see the Appendix),
in the limit $k_c \gg 1$ 
\be
{1 \over p_u} = \left[{1+{1 \over 2 \pmax-1} \over (e^2-1)}\right] e^{2 k_c}.
\ee
The behavior of $1/p_u$ with decreasing $\les$ is then due to
a power-law divergence of the critical height $k_c$.
Since the height and the width of the critical nucleus scale in the
same way~\cite{note_k-Lambda} 
it is then possible to write $1/p_u \sim e^{a(m) \Lambda_c}$.

\section{The effect of thermal detachment}
\label{sec_thermal}

We now analyize how the previous picture of the breakdown
of the metastable state is modified by the possibility that
atoms detach from steps or islands with rate $R_d$.
Figures~\ref{Fig_tauins}, ~\ref{Fig_pu}, and \ref{Fig_Lambda}
show the effect of $R_d$ 
on the instability time and on the critical width $\Lambda_c$,
respectively.
At first sight these results might seem contradictory.
While the critical nucleus width does not change appreciably
when $R_d$ grows from 0 to 70 (Fig.~\ref{Fig_Lambda}), $1/p_u$
increases by a factor larger than $10^3$ (Fig.~\ref{Fig_pu}).

In Sec.~\ref{sec_mound-width} we related $\Lambda_c$ to $L_c$, the
minimal length for the onset of the (linear) instability on a
singular surface.
A numerical check shows that also $L_c(\les)$ does not change
when thermal detachment is allowed:
we can therefore conclude that Eq.~(\ref{Lambda_c}) remains valid
also in the presence of thermal detachment.
From the continuum theory (Sec.~\ref{sec_cahn-hilliard}),
$L_c \sim \sqrt{K/\les}$, where the coefficient $K$ contains several
contributions, one of which
is due to thermal detachment, $K = K_0 + K_{ther}$.
The fact that $L_c$ is independent of $R_d$ implies that for $R_d\le 70$
the thermal contribution $K_{ther}$ is negligible.

\begin{figure}
\includegraphics[angle=0,width=8cm,clip]{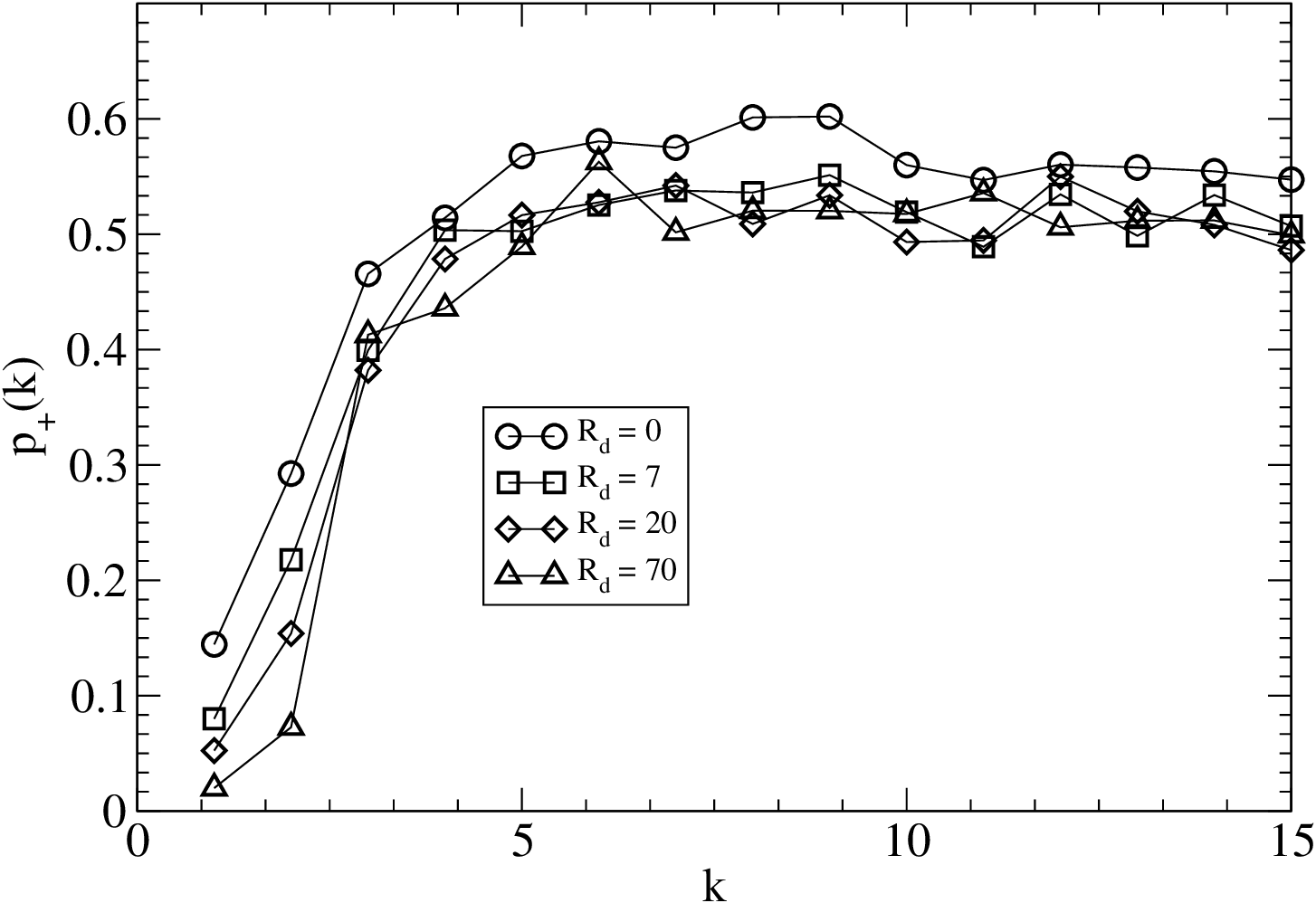}
\caption{The probability $p_+(k)$ that a mound of height $k$ eventually
turns into a mound of height $k+1$, rather than decaying to a mound
of height $k-1$. Results are for $m=0.2$ and $\les=500$.}
\label{Fig_pofk_det}
\end{figure}

To clarify why thermal detachment has instead dramatic effects
on $1/p_u$, we have computed, exactly as for $R_d=0$, the
transition probabilities $p_+(k)$ for the simplified random walk
picture (Fig.~\ref{Fig_pofk_det}).
The presence of thermal detachment leaves the shape of $p_+(k)$
qualitatively the same.
The value of the threshold $k_c$ (the critical nucleus height)
does not change much with $R_d$, in agreement with the result
$\Lambda_c$: the critical nucleus does not practically depend on $R_d$.
However, for the same $k$, $p_+(k)$ is reduced
as $R_d$ grows, in particular for small $k$.
This tendency has a simple interpretation in terms of a
microscopic process induced by thermal detachment: 
the decay of the freshly nucleated dimers on top of mounds.
While for $R_d=0$ a dimer on top of a mound is stable and can only
grow by incorporating incoming adatoms, for $R_d>0$ a dimer can
split in two adatoms, which may diffuse and be incorporated by
surrounding steps, restoring the situation with no dimer.
This decay reduces the probability $p_+(k)$ that a mound grows higher.
A smaller $p_+(k)$ clearly implies an increased $1/p_u$.
Conversely, when a dimer splits during the growth of a
high-symmetry surface, adatoms are not lost because there are no
preexistent steps which capture them: hence such a splitting has
no dramatic effect, leaving $L_c$ (and hence on $\Lambda_c$) unchanged.
This is the reason why switching $R_d$ on is relevant for $\tins$
and not for $\Lambda_c$.

The reduction of $p_+(k)$ shown in Fig.~\ref{Fig_pofk_det} does not
tell the whole story about the effect of thermal detachment.
If we compare the value of $1/p_u$ determined numerically for the
{\em uncorrelated} random walk with transition probabilities $p_+(k)$
with the same quantity measured from the full KMC simulations
(Fig.~\ref{Fig_pu_comp}), we realize that the results do not agree
and the mismatch grows with $R_d$.
The smaller $p_+(k)$ is not enough to explain the instability time measured
in KMC simulations:
the assumption of uncorrelated steps is wrong when $R_d\ne 0$.
The origin of the correlation is another subtle but relevant effect
of the decay of dimers.
When a dimer is formed on the top terrace of a mound ($k \to k+1$),
owing to thermal detachment it is very easy that the dimer will soon
decay so that $k+1$ goes back $k$. This increased chance that a growth
event is immediately followed by a decrease of $k$ invalidates the
assumption of uncorrelated steps.
Therefore, the uncorrelated random walk picture is no more useful
for determining quantitatively $1/p_u$ and from it the instability time.

\section{Conclusions}
\label{sec_conc}

In a nutshell, we have investigated the breakdown
of the metastable step-flow for epitaxially grown vicinal surfaces.
We have performed Kinetic Monte Carlo simulations showing that
metastability holds for any strength of the ES barrier: the time needed
for destabilization to occur diverges exponentially when the barrier
goes to zero, both in the presence and in the absence of thermal
detachment.
Previous atomistic and continuum approaches to the problem do not
reproduce this phenomenology.
The crucial event determining the end of the metastable state is 
the formation, via a rare fluctuation, of a critical nucleus whose
size diverges as the barrier goes to zero.
The dependence of the instability time on the parameters of the system
can be summarized as
\be
\tins(L,m,\les,R_d) \approx
{\tdim(m,\les,R_d)\over (m L)} \
e^{[a(m,R_d) L_c(\les)]},
\ee
where $\tdim$ is the time to nucleate a dimer on a vicinal terrace and 
$L_c(\les)$ is the length associated to the linear instability on a
singular surface. $\tdim$ has a well-known analytical form\cite{Politi03}
(at least for $R_d=0$) and $L_c=\sqrt{K\over\les}$. The only quantity which
has not been determined is the prefactor $a$ in the
exponent. The $\les$-dependence in $\tdim$ is weak and
negligible for small $\les$.

The results presented fully clarify how the destabilization of
step-flow comes about and also provide a semi-quantitative picture.
The analytical interpretation would be completed by a direct computation
of the probabilities $p_+(k)$ based on the microscopic dynamics.
This remains a challenge for future work.

Another intriguing open question has to do with the difficulties of the
continuum approach. In Sec.~\ref{sec_cahn-hilliard} we have shown 
that the minimal model with $J=Ku_{xx}+\alpha\jes(u)$ is completely unable to
explain the phenomenology of the metastability process.
A likely explanation for the failure of the continuum theory
is that the current $J$ considered is {\it too} simple: it is well known
that terms breaking the $z\to -z$ symmetry may be relevant~\cite{review},
and a look at profiles from KMC simulations indicate that they play some role.
These symmetry-breaking terms usually make the equation of motion
nonderivable from a potential so that energetic arguments as those
presented in Sec.~\ref{sec_cahn-hilliard} are not applicable.
An analytical approach to the metastability of generalized Cahn-Hilliard
equations with symmetry breaking terms is still to be found.
But solving this problem may not be enough.
Because of such additional terms, non-analytical regions could appear
in the surface profile, in particular in the surroundings of minima
where mounds merge. If this occurs a continuum approach may not be
altogether applicable.

Despite these possible objections to the use of the minimal model
(\ref{eq_min}) it is important to remind that such a model is instead
appropriate to describe at least qualitatively the linear and the nonlinear
regimes of the destabilization of a singular surface.
Why the model is not appropriate for the understanding the decay of the
metastable regime is, in this light, a puzzle.

Finally, some considerations on two-dimensional systems.
The argument given in Sec.~\ref{sec_crit-nuc}, based on the hypothesis that
critical mounds have constant height, can be easily extended to $d=2$
and still provides a threshold.
Within our approach, we can 
write the analogue of Eq.~(\ref{eq_nucl-pu}),
\be
\tins(L_\parallel,L_\perp,\les) =
{\tdim(\ell)\over L_\perp (L_\parallel/\ell)} {1\over p_u} ,
\ee
where $L_\parallel$ and $L_\perp$ are the linear size of the system
in the direction parallel and perpendicular to the slope, respectively.
$\tdim (\ell)$ is the time to form a dimer on a terrace of size
$\ell$ in the parallel direction per unit length in the perpendicular
direction.
Again, $p_u(\les)$ is the probability that a dimer grows up to the critical 
size instead of being absorbed.
Assuming that the form of the probabilities $p_+(k)$ will be similar
to the one-dimensional case, it is reasonable to expect
$1/p_u \sim \exp(k_c)$ also in this case.
Since $k_c$ scales as $\Lambda_c$ in $d=2$ as well, 
we conjecture that also in two dimensions the instability time
diverges as $\exp(a/\les^{1/2})$ when the ES barrier goes to zero.

\appendix

\section{Random walk model}

In this Appendix we compute analytically the probability $p_u$
that a freshly created dimer evolves into an unstable mound instead of
being reabsorbed, given the probabilities $p_+(k)$ [$1-p_+(k)$] of the
process $k \to k+1$ [$k \to k-1$].
The problem is equivalent to the computation, for a random walk with
absorbing boundaries in 0 and $N \to \infty$, of the exit probability in $N$,
with the walker starting initially in $k=1$.

Let us call $\e(x)$ the exit probability if the walker is deposited in $x$.
This quantity obeys the recurrence relation~\cite{Rednerbook}

\be
\e(x) = p_+(x) \e(x+1) + (1-p_+(x)) \e(x-1) ,
\ee
with boundary conditions $\e(0)=0$ and $\e(N \to \infty)=1$.

In the continuum limit
\be
0 = a(x) \e'(x) + {1 \over 2} \e''(x) ,
\ee
with $a(x) = 2 p_+(x)-1$.
Therefore we must solve a stationary convection-diffusion equation
with space-dependent drift.
The equation can be solved by separation of variables yielding
\be
\e(x) = C_1 + C_0 \int_0^x dx' e^{-2 f(x')} ,
\ee
with $f(x') = \int_0^{x'} a(x'') dx''$.

To proceed further we specify the form of $p_+(k)$, in agreement
with Fig.~\ref{Fig_pofk},
\be
p_+(k) = \left\{
\begin{array}{ll}
p_0 e^k & k<k_c \\
\pmax     & k>k_c
\end{array}
\right.
\ee
with $p_0 e^{k_c} = \pmax$.
Using this form we have, for $x<k_c$,
\be
f(x) \equiv f_1(x) = 2 p_0 e^x - x ,
\ee
so that, for $x<k_c$,
\bea
&& \e(x) \equiv \e_1(x) \\
&& = C_1 +
C_0 {(1+4 p_0) e^{4 p_0} - (1+4 p_0 e^x) e^{-4 p_0 e^x} \over 16 p_0^2} ,
\nonumber
\eea
while for $x>k_c$
\be
\e(x) \equiv \e_2(x) = \e_1(k_c) + C_0 {e^{-2 f_1(k_c)} \over 2(2 \pmax -1)} .
\ee

The boundary condition $\e(0)=0$ sets $C_1=0$.
The second boundary condition determines the value of $C_0$
\be
{1 \over C_0} =
{(1+4 p_0) e^{4 p_0} - (1+4 p_0 e^x) e^{-4 p_0 e^x} \over 16 p_0^2} + 
{e^{-2 f_1(k_c)} \over 2(2 \pmax -1)} .
\ee

The quantity we are interested in is
$1/p_u = 1/\e_1(x=1)$.
In the limit  $k_c \gg 1$, which implies $p_0 \to 0$,
and $e^{-2 f_1(k_c)} \approx e^{2 k_c}$, we find
\be
{1 \over p_u} = \left[ {1+{1 \over 2 \pmax-1} \over (e^2-1)}\right]e^{2 k_c}.
\ee


\begin{thebibliography}{99}
\bibitem{note_meander}
Additional processes not considered here, e.g. diffusion along edges,
might stabilize the step profile.
\bibitem{note_KK}
Relaxing this assumption and performing a more precise calculation
does not modify the result.
\bibitem{note_k-Lambda}
The slope of a mound may increase with its size rather than
remaining constant, but this has a negligible effect in the
range of values of $\Lambda_c$ and $k_c$ considered here:
these two quantities can safely be assumed to scale in the
same way, $k_c\sim\Lambda_c$.
\end{thebibliography}
\end{document}